\documentclass[runningheads,a4paper]{llncs}

\usepackage{amssymb}
\usepackage{graphicx}

\usepackage{amsmath}
\usepackage{subfig}
\usepackage{tikz}
\usetikzlibrary{arrows,snakes,backgrounds}
\usetikzlibrary{automata,positioning}
\usepackage{calc}
\usepackage{ifthen}
\usepackage{algorithm,algpseudocode}
\usepackage{url}
\usepackage{multirow}
\usepackage{rotating}
\usepackage{tipa}

\usepackage{url}
\urldef{\mailsa}\path|{alfred.hofmann, ursula.barth, ingrid.haas, frank.holzwarth,|
\urldef{\mailsb}\path|anna.kramer, leonie.kunz, christine.reiss, nicole.sator,|
\urldef{\mailsc}\path|erika.siebert-cole, peter.strasser, lncs}@springer.com|    
\newcommand{\keywords}[1]{\par\addvspace\baselineskip
\noindent\keywordname\enspace\ignorespaces#1}

\begin{document}

\mainmatter  

\title{A Probabilistic Translation Method for Dictionary-based Cross-lingual Information Retrieval in Agglutinative Languages}

\titlerunning{A Probabilistic Translation Method for Dictionary-based Cross-lingual Information Retrieval in Agglutinative Languages}

%
%
\author{Javid Dadashkarimi%
\and Azadeh Shakery\and Heshaam Faili}

\urldef{\mailsa}\path|{dadashkarimi, shakery, hfaili}@ut.ac.ir|    

\institute{School of Electrical and Computer Engineering,\\
College of Engineering,\\
University of Tehran, Tehran, Iran. \\
\mailsa\\}

%
%

\toctitle{Minimum Edit Support Candidates Model}
\tocauthor{Authors' Instructions}
\maketitle

\begin{abstract}
Translation ambiguity, out of vocabulary words and missing some translations in bilingual dictionaries make dictionary-based Cross-language Information Retrieval (CLIR) a challenging task. Moreover, in agglutinative languages which do not have reliable stemmers, missing various lexical formations in bilingual dictionaries degrades CLIR performance. This paper aims to introduce a probabilistic translation model to solve the ambiguity problem, and also to provide most likely formations of a dictionary candidate. We propose Minimum Edit Support Candidates (MESC) method that exploits a monolingual corpus and a bilingual dictionary to translate users' native language queries to documents' language. Our experiments show that the proposed method outperforms state-of-the-art dictionary-based English-Persian CLIR. \footnote{\it{In Proceedings of Third Computational Linguistic Conference in Sharif University of Technology, Nov 2014}}
\end{abstract}
\section{Introduction}
Languages are shared progressively in the World Wide Web. As a result, there is relatively remarkable research on Cross-language Information Retrieval (CLIR) to extract the information in such a large multilingual data. CLIR tasks mostly focus on retrieving documents in a language different from users' native language and presenting the documents in a ranked list based on their relevance to users' queries. Retrieving documents in this way can be done by employing the following approaches. 1- Translating queries to the target language, 2- Translating documents to the source language, or 3- Mapping documents and queries to a third language \cite{kishida2005,pirkola98,nie2010,levow2005,donnla2005,young:1994}. Since document translation is a time consuming and costly task, query translation is preferred. Query translation can be done by one of the following methods: 1-~Using bilingual corpora and extracting a probabilistic dictionary, 2-~Translation by using a Machine Translators (MT), or 3-~Exploiting the bilingual machine readable dictionaries. Independence assumption of query terms in retrieval methods, difficulty of creating reliable MTs, and scarcity of the aligned corpora in many language pairs, make dictionary-based translation an available and straightforward solution in CLIR tasks.\\
Dictionaries provide a list of translations for each query term. According to \cite{donnla2005}, ambiguity in translation and swamping effect\footnote{retrieving irrelevant documents by selecting irrelevant candidates.} are the most important challenges of dictionary-based methods. This paper aims to introduce Minimum Edit Support Candidates (MESC) method, a probabilistic model to overcome these challenges specifically in morphologically rich languages. In morphologically rich languages, candidates have different formations according to their parts of speech while dictionaries cannot provide all of these translations. Indeed, adding most similar words to dictionary candidates and employing a probabilistic candidate selection model, could substantially improve the CLIR task. Indeed, the proposed MESC exploits a monolingual corpora to extract different formations of the dictionary candidates. Furthermore the proposed algorithm considers other query terms to generate most probable formations. In final step, MESC builds its translation model based on candidates' bigram probabilities. Additionally it uses a simple rule-based transliterator to overcome the Out Of Vocabulary (OOV) problem.\\
Persian is an example of highly inflected languages and there is not an effective stemmer in Persian. Experiments are specifically centered on English-Persian CLIR task. Queries are in English and documents' language is Persian. We use INQUERY retrieval system \cite{pirkola98}, rank-based methods and monolingual runs as our baselines. Experimental results show that MESC outperform the previous dictionary-based CLIR approaches.\\
The rest of the paper is organized as follows: Section~\ref{PreviousWorks} reviews previous works on dictionary-based CLIR. In Section~\ref{MESC} we explain the MESC algorithm in details. Comparing MESC with the Pirkola's structured query system and other rank-based methods is discussed in Section~\ref{Experiments}. We bring future works and conclude the paper in Section~\ref{Conclusion}.
\section{Previous Works}
\label{PreviousWorks}
Bilingual dictionaries are truthful resources for compound word detection \cite{chen2001} and query expansion \cite{cao2007}. Ambiguity resolution is also done by Maximum Coherence Model (MCM) \cite{liu2005} or graph-based models \cite{Zhou:2008:HTE:1362782.1362784}. MCM concerns with translation consistency within context. Mutual Information (MI) between dictionary translations be used as a measure of similarity in MCM. Phrase translation, specifically in morphologically rich languages is a main difference between MCM and MESC. As a similar analysis, graph-based models focus on correlations between translations. Authority score and hub score are translation scores in the graph-based approach. \\
Azarbonyad \emph{et al.} employ ranks of candidates to query translation \cite{hosein2012,azarbonyad2013}. Top $N$ candidates after parameter tuning are selected as final translations. Swamping effect is the most important side effect of the rank-based methods. On the other hand Hashemi \emph{et al.} utilize a bilingual dictionary and a comparable corpus to build their Term Association Network (TAN) \cite{hashemi2011}. Translations and relevant terms are extracted after creating TAN. Subject dependency of the exploited bilingual corpora is the main drawback of TAN. Nic \emph{et al.} compute degree of ambiguity for each candidate and eliminate ambiguous translations \cite{donnla2005}. Tuning the ambiguity threshold is aim of the Nic \emph{et al.}'s approach. Since ambiguity is a context dependent problem and Nic \emph{et al.}' approach is a dictionary characteristic-based solution, the degree of ambiguity is not an effective solution. Sense disambiguation upon context terms proposed by Kishida \cite{Kishida01082009}. Despite Kishida's attempts to disambiguate between multiple translations, he uses a sentence aligned corpora. 
\\Pirkola introduces INQUERY, a probabilistic retrieval system \cite{pirkola98,pirkola2001} and Oard \cite{oard2000} presents an overlapping character bigram-based approach, which is suitable for Chinese character re-segmentation problem. Regardless of these methods' aims to solve the ambiguity problem, crucial equivalent problem\cite{donnla2005}, and phrase translation problem, they lack accuracy: missing different formations of a candidate in a morphologically rich language, and missing powerful stemmers in such a language, make these methods fail to match accurate formations. 
\\Finding most similar formations in MESC is strongly related to error correction approaches specifically in previous Persian language studies \cite{Ehsan2013,Miangah2014}.

\section{Dictionary-based CLIR and MESC}
\label{MESC}
Previous dictionary-based CLIR studies employ bilingual dictionaries to translate queries to retrieve documents in a different language from queries' language. Dictionary-based algorithms suffer from ambiguity in candidate selection. Moreover, dictionaries cannot provide all formations of a candidate. For instance, in \textit{`World Cup'} query, we expect CLIR algorithm to generate \textit{`j$\hat{a}$m jh$\hat{a}$ni'} as the correct translation. But 1:\textit{`jh$\hat{a}$n'}, 2:\textit{`giti'}, 3:\textit{`dni$\hat{a}$'}, and 4:\textit{`$\hat{a}$lm'} are only the provided translations in a bilingual English-Persian dictionary for \textit{`World'} and 1:\textit{`fnj$\hat{a}$n'}, 2:\textit{`j$\hat{a}$m'}, and 3:\textit{`pi$\hat{a}$lh'} for \textit{`Cup'}. Stemming problem in highly inflected languages as well as ambiguity problem make the previous dictionary-based CLIR methods to have degraded performances. This paper aims to overcome the mentioned challenges in CLIR task. The view presented in the proposed Minimum Edit Support Candidates (MESC) is to add similar terms based on their forms to dictionary candidate lists. Similar terms which have co-occurrences with candidates belonging to other query terms are only selected. For example \textit{`jh$\hat{a}$ni'} is not only similar to \textit{`jh$\hat{a}$n'} but also it is co-occurred with \textit{`j$\hat{a}$m'}, the second translation of \textit{`Cup'} in bilingual dictionaries.\\
Minimum Edit Distance (MED) is an algorithm to find minimum number of steps transforming one string into another, in terms of insertion, deletion and substitution. There are also other versions of the algorithm, but the Levenshtein algorithm is a simple version that assumes the weights of all operations to be equal \cite{levenshtein1966bcc,jurafsky2009speech}. We apply the Levenshtein algorithm to find similar terms to dictionary candidates.\\
Despite the support lists' attempts to solve the crucial equivalent selection problem, selecting most relevant translation, it has a drawback: Adding such derivational forms may cause selecting wrong translations due to generating irrelevant lexicons by MESC. Moreover it is important to decide which candidates in $\mathbf{c}_i$ and $\mathbf{s}^{lev}_i$ is the best translation according to the context. In fact, ignoring noisy support candidates and solving the ambiguity problem are other challenges of MESC. 
\\We present some notations in Section~\ref{notations}. The OOV word problem is discussed in Section~\ref{OOV}. Extracting the support candidates and details of MESC's translation model are presented in Section~\ref{ESC} and Section~\ref{MESCM} respectively.\\

\subsection{Notations}
\label{notations}
In our CLIR task, there is a set of documents in target language which is called $\mathbf{D}^t=\{\mathbf{d}_1,\mathbf{d}_2,..,\mathbf{d}_m\}$ and a set of source language query terms $\mathbf{Q}^s=\{q^s_1,q^s_2,..,q^s_m\}$ where $q^s_i$ is the $i$-th query term. Each query term has a list of dictionary candidates in $\mathbf{C}^t=\{\mathbf{c}_1,\mathbf{c}_2,..,\mathbf{c}_m\}$. Furthermore $\mathbf{V}^t=\{{v}_1,{v}_2,..,{v}_n\}$ is the vocabulary set in target language and $\mathbf{A}=[a_{x,y}]_{|\mathbf{V}^t|\times |\mathbf{V}^t|}$ indicates the adjacency matrix. $a_{x,y}$ equals to one if two terms co-occurred within a specified window $w$ and equals to zero otherwise.
Similar terms to the dictionary candidates are extracted from a monolingual corpus and make secondary lists or the support lists for the query terms. We have a support set $\mathbf{S}^t_{lev}=\{\mathbf{s}^{lev}_1,\mathbf{s}^{lev}_2,..,\mathbf{s}^{lev}_m\}$, where $\mathbf{s}^{lev}_i$ stands for a list of some $v_j \in \mathbf{V}^t$ which its minimum edit distance from a candidate in $\mathbf{c}_i$ is equals to one or two. Whose every query term $q_i$ has two lists of translation candidates $\mathbf{c}_i$ and $\mathbf{s}^{lev}_i$ and other forms of a lexicon in $\mathbf{c}_i$ can appear in the latter list. Finally $\mathbf{Q}^t=\{q^t_1,q^t_2,..,q^t_m\}$ is a set of translated query terms.

\vspace{-0.2cm}
\subsection{Out Of Vocabulary}
\label{OOV}
Plural nouns, different formations of a verb, proper nouns, and phrases form main parts of dictionary-based CLIR challenges. Regardless of possibility of determining stems of plural query terms by applying simple algorithms, plural form of a candidate may be produced by MESC if its minimum edit distance from its singular form is equals to one or two. For an example, in query \textit{`Iran Football Coaches'}, the MESC algorithm generates \textit{`mrbi$\hat{a}$n futb$\hat{a}$l ir$\hat{a}$n'} which \textit{`mrbi$\hat{a}$n'} is generated due to the reason of having two edit distances with its singular form \textit{`mrbi'}. Proper nouns are transliterated using a probabilistic rule-based transliterator. Replacing all consonant letters, generating all possible replacements of vowels and pushing all of them in query term's support candidate list is the first step in proper noun translation. Secondly, the MESC algorithm aims to select the most probable formation based on its bigram probability with other query terms.\\
Fig.~\ref{proportions} presents statistics of the sources of errors in query translations in topics of CLEF 2008 and 2009. Topics in CLEF 2009 consist of more ambiguous queries. Existing a great number of phrases and multiple phrases within a query in topics of CLEF 2009 make CLIR a more challenging task in such topics.
\newcommand{\slice}[4]{
  \pgfmathparse{0.5*#1+0.5*#2}
  \let\midangle\pgfmathresult

  \draw[thick,fill=black!10] (0,0) -- (#1:1) arc (#1:#2:1) -- cycle;

  \node[label=\midangle:#4] at (\midangle:1) {};

  \pgfmathparse{min((#2-#1-10)/110*(-0.3),0)}
  \let\temp\pgfmathresult
  \pgfmathparse{max(\temp,-0.5) + 0.8}
  \let\innerpos\pgfmathresult
  \node at (\midangle:\innerpos) {#3};
}

\definecolor{myblue}{HTML}{E8E8E8}
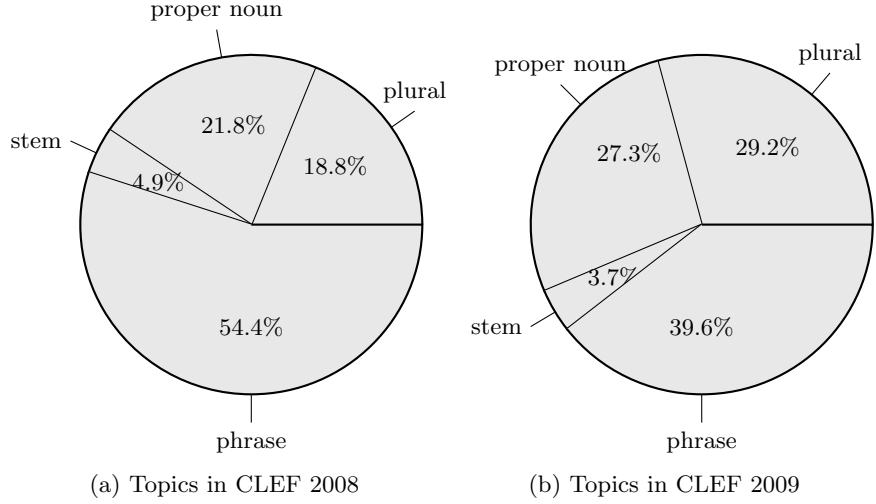
\begin{figure}[t]
	\subfloat[Topics in CLEF 2008]{
	\begin{tikzpicture}[scale=0.75]
	
	    \foreach \start/\end/\middle/\percent/\anchor/\name/\smeter in {
	      0/68/35/18.8/above/plural/,
	      68/146/100/21.8/above/proper noun/,
	      146/162/155/4.9/left/stem/,
	      162/360/270/54.4/below/phrase/}
	  {
	    \draw[fill=myblue, thick] (0,0) -- (\end:3cm) arc (\end:\start:3cm)
	      node at (\middle:1.8cm) {\percent \%};
	    \draw (\middle:3cm) -- (\middle:3.5cm) node[\anchor] {\name \ \smeter};
	  };

	\end{tikzpicture}}
    \subfloat[Topics in CLEF 2009]{
	\begin{tikzpicture}[scale=0.75]
	
	    \foreach \start/\end/\middle/\percent/\anchor/\name/\smeter in {
	      0/105/50/29.2/above/plural/,
	      105/203/135/27.3/above/proper noun/,
	      203/218/211/3.7/left/stem/,
	      218/360/270/39.6/below/phrase/}
	  {
	    \draw[fill=myblue, thick] (0,0) -- (\end:3cm) arc (\end:\start:3cm)
	      node at (\middle:1.8cm) {\percent \%};
	    \draw (\middle:3cm) -- (\middle:3.5cm) node[\anchor] {\name \ \smeter};
	  };

	\end{tikzpicture}}
		\caption{Portions of out of vocabulary words in topics of CLEF 2008 and 2009.}
		\label{proportions}
\end{figure}
\vspace{-0.25cm}

\subsection{Extracting Support Candidates}
\label{ESC}
The forgoing discussions imply that support candidates are arbitrary options to be chosen as a translation of a query term. In order to extract these words, the adjacency matrix $\mathbf{A}$ is firstly filled with binary numbers indicating closeness of terms in a target language collection. Accordingly, for a dictionary candidate $c_{i,j}$, the support candidate $s^{lev}_{i,j^{\prime}}$ is some $v\in \mathbf{V}^t$ that not only has one or two minimum edit distances from $c_{i,j}$ but also it has a non-zero co-occurrence value with at least a term in other query terms' candidates. In brief, support candidate list for the $i$-th query term is:

\begin{equation}
\label{eq:support}
\begin{array}{ll}
\quad\mathbf{s}^{lev}_{i}=\big\{v\in \mathbf{V}^t| & \exists c_{i,j} \in \mathbf{c}_i ~\big(\exists c_{i\prime,j\prime} \in \mathbf{c}_{i\prime}~\big(i\neq i^\prime ~\wedge~  \\&  a_{v,c_{i\prime,j\prime}} =1 ~\wedge~ 1\leq \text{MED}(v,c_{i,j})~\leq 2\big)\big)\big\} .
\end{array}
\end{equation}

\subsection{Minimum Edit Support Candidates Model}
\label{MESCM}

On balance, each query term has two lists of candidates and selecting the best candidate is aim of the current discussion.
MESC defines two sets of probability lists: 
\begin{enumerate}
\item 
$\mathbf{P}^c=\{\mathbf{p}^{c}_1,\mathbf{p}^{c}_2,..,\mathbf{p}^{c}_m\}$ denotes a set of lists like $\mathbf{p}^{c}_i$ that contains probabilities of all candidates in $\mathbf{c}_i$ conditioned on observing the $q_i$ (to be exact, $p^c_{i,j}=P(c_{i,j};q^s_i)$ which $c_{i,j}$ indicates the $j$-th candidate of $\mathbf{c}_i$).
\item Similarly $\mathbf{P}^{s_{lev}}=\{\mathbf{p}^{s_{lev}}_1,\mathbf{p}^{s_{lev}}_2,..,\mathbf{p}^{s_{lev}}_m\}$ consists of lists of probabilities corresponding each query term's support candidates conditioned on observing its query term (more precisely $p^{s_{lev}}_{i,j}=P(s^{lev}_{i,j};q^s_i)$ that $s^{lev}_{i,j}$ points to the $j$-th candidate of the support list $\mathbf{s}^{lev}_i$). 
\end{enumerate} 
In short:
\begin{equation*} 
\begin{array}{lcl c lcl}
\mathbf{p}^c_i & = & [p^c_{i,j}]_{1\times |\mathbf{c}_i|} &\hspace{10mm}&
\mathbf{p}^{s_{lev}}_i & = & [p^{s_{lev}}_{i,j}]_{1\times |\mathbf{s}^{lev}_i|}  \end{array} .
\end{equation*}

The probability of a candidate after observing its query term can be shown as:

\begin{equation}
\label{eq5}
\begin{array}{lll}
P(c_{i,j};q^s_i)=&\sum\limits_{i^\prime \neq i}^{}\Big(&\sum\limits_{j^\prime=1}^{|\mathbf{c}_{i^\prime}|}P(c_{i,j}|c_{i^\prime,j^\prime};q^s_i)P(c_{i^\prime,j^\prime};q^s_{i})+\\
&&\sum\limits_{j^\prime=1}^{|\mathbf{s}^{lev}_{i^\prime}|}P(c_{i,j}|s^{lev}_{i^\prime,j^\prime};q^s_i)P(s^{lev}_{i^\prime,j^\prime};q^s_{i})\Big) .\\
\end{array}
\end{equation}

As a simplification assumption, translation candidates are assumed independent from other source query terms (i.e. , $P(c_{i^\prime,j^\prime};q_i)\approx P(c_{i^\prime,j^\prime})$). Furthermore we can estimate $P(c_{i,j}|c_{i^\prime,j^\prime};q_i)$ using bigram probability of the candidates. In more detail we have:
\begin{equation}
\begin{array}{ll}
P(c_{i,j}|c_{i^\prime,j^\prime};q^s_i)P(c_{i^\prime,j^\prime};q^s_i)&
\approx P(c_{i,j}|c_{i^\prime,j^\prime})P(c_{i^\prime,j^\prime})\\
&= P(c_{i,j},c_{i^\prime,j^\prime}) .
\end{array}
\end{equation} 
Similarly:
\begin{equation}
\begin{array}{ll}
P(c_{i,j}|s^{lev}_{i^\prime,j^\prime};q^s_i)P(s^{lev}_{i^\prime,j^\prime};q^s_{i})&
\approx P(c_{i,j},s^{lev}_{i^\prime,j^\prime})
\end{array}
\end{equation} 

 If we define $p^c_{i,j}=P(c_{i,j};q^s_i)$ the Equation \ref{eq5} can be represented as follow:
 
\begin{equation}
\label{eq6}
p^c_{i,j}=\sum\limits_{i^\prime \neq i}^{}\Big(\sum\limits_{j^\prime=1}^{|\mathbf{c}_{i^\prime}|}P(c_{i,j},c_{i^\prime,j^\prime})+\sum\limits_{j^\prime=1}^{|\mathbf{s}^{lev}_{i^\prime}|}P(c_{i,j},s^{lev}_{i^\prime,j^\prime})\Big) .
\end{equation} 
  
As be stated, every dictionary candidate's probability depends on terms in the other query terms' candidate lists. But in this case the probability of a support candidate is obtained just by comparing with dictionary candidates to prevent adding noisy probabilities. Indeed the Levenshtein algorithm may produce an irrelevant support candidate which have a high bigram probability with another irrelevant support candidate. As a result a support candidate probability equals:

\begin{equation}
\label{eq7}
\begin{array}{lll}
p^{s_{lev}}_{i,j}=&\sum\limits_{i\prime\neq i}^{}&\sum\limits_{j\prime=1}^{|\mathbf{c}_{i\prime}|}P(s^{lev}_{i,j},c_{i\prime,j\prime}) .
\end{array}
\end{equation} 

Final normalization consideration can be stated as follows:
\begin{equation}
\label{eq9}
\sum\limits_{j=1}^{|\mathbf{c}_i|}p^c_{i,j} + \sum\limits_{j=1}^{|\mathbf{s}^{lev}_i|}p^{s_{lev}}_{i,j}=1 .
\end{equation}
 
The Equation \ref{eq9} points to probability distribution condition for translations of the $q^s_i$. It seems evident that Equation \ref{eq9} forces $p$ to distribute probability over the candidates in dictionary and the support list. Any other translation that is not presented in such lists gets zero probability. Indeed, coverage of the dictionary plays an important role in the proposed model.

\subsection{Candidate Selection}
\vspace{-0.15cm}
Finally, if we define $\mathbf{T}$ a translation candidate list, and $\mathbf{P}^T$ a corresponding probability vector, $q^t_i$, the final translation of $q^s_i$, is a candidate with highest probability:\\
\begin{equation*} 
\begin{array}{lcl c lcl}
\mathbf{T} & = & \begin{pmatrix}
    \mathbf{c}_i \\
    ~~\mathbf{s}^{lev}_i
\end{pmatrix} &\hspace{10mm}&
\mathbf{P}^T & = &  \begin{pmatrix}
   \mathbf{P}^c_i \\
   ~~~\mathbf{P}^{s_{lev}}_i
  \end{pmatrix}  \end{array}
\end{equation*}

\begin{equation}
q^t_i=\underset{T_{i,j}} {\mathrm{arg~max}} {~p^T_{i,j}}.
\end{equation}

\section{Experiments}
\vspace{-0.15cm}
\label{Experiments}
This section attempts to investigate the validity of the proposed algorithm. Indeed implementation the of MESC algorithm in topics of CLEF 2008 and CLEF 2009 and Hamshahri\cite{hamshahri2009} collection is discussed in the current section.

\subsection{Experiment Setup}
\subsubsection{Data Collection}
Hamshahri is a Persian document collection with 166,774 documents whose average document length equals to 225 terms. This collection has been used as a resource for our retrieval task and computing unigram and bigram probabilities or joint probabilities to be exact. We can use any other large monolingual collection in target language to compute such probabilities.
\vspace{-0.5cm}
\subsubsection{Tools and Toolkits}
Lemur toolkit \footnote{http://www.lemurproject.org} is our retrieval tool and Okapi is selected as a retrieval model \cite{Robertson94}. Pseudo relevance feedback has been done as a query expansion phase for all monolingual retrieval runs, the proposed algorithm, and previous works. Probabilities are extracted using SRILM toolkit\cite{Stolcke02}.
\vspace{-0.5cm}
\subsubsection{Bilingual Dictionary}
Regarding comparing the CLIR performance of MESC with the previous methods, proposed algorithm is applied on title parts of the topics in CLEF 2008 and CLEF 2009. Each of them consists of 50 English queries. In addition to Hamshahri we exploit three machine readable bilingual dictionaries: 
\begin{enumerate}
\item Aryanpour\footnote{http://www.aryanpour.com} bilingual dictionary which has been used in most of the previous dictionary-based English-Persian CLIR research \cite{hosein2012,hashemi2014}. 
\item Dictionary of Google\footnote{http://translate.google.com} and not using its MT. 
\item Faraazin\footnote{http://www.faraazin.ir} and using its available bilingual dictionary.
\end{enumerate}

Table~\ref{scaleTable} shows average number of candidates for each entry of a dictionary or dictionary scales \cite{donnla2005}. Aryanpour has almost largest candidate lists. Coverage of dictionaries over topics in CLEF 2008 and CLEF 2009 are equal and the dictionaries' characteristics differ according to their scales and rankings.

\begin{table}[!h]
\caption{Average number of candidates for each dictionary entry.}   
\begin{center}
    \begin{tabular}{p{4cm} p{3cm} p{3cm} p{1cm}}
    \hline\noalign{\smallskip}
    Characteristic & Aryanpour & Google & Faraazin \\
    \noalign{\smallskip}
    \hline
    \noalign{\smallskip}
   Dictionary Scale &	3.83 &	3.36	& 3.30 \\ 
   Candidates Variance & 2.83  & 2.35  & 2.48 \\ \hline
    \end{tabular}
    \label{scaleTable}
\end{center} 
\end{table}

\subsection{Persian Monolingual Retrieval}
Monolingual Persian retrieval is our CLIR task's baseline. Table~\ref{mono} shows results of the monolingual retrieval runs on the topics in CLEF 2008 and CLEF 2009.

\vspace{-0.2cm}
\begin{table}[!h]
\caption{Monolingual retrieval results.}   
\begin{center}
    \begin{tabular}{p{4cm} p{3cm} p{3cm} p{1.5cm}}
    \hline\noalign{\smallskip}
     Queries       & MAP & Prec@5 & Prec@10 \\
    \noalign{\smallskip}
    \hline
    \noalign{\smallskip}
   CLEF 2008 &	0.4449 & 0.7040	& 0.6720 \\ 
   CLEF 2009 & 0.4070  & 0.6000 & 0.5980 \\ \hline
    \end{tabular}
    \label{mono}
\end{center} 
\end{table}
\vspace{-0.2cm}
\subsection{Top Ranked Candidates vs Pirkola's Structured Query}
Most studies such as \cite{donnla2005,hashemi2014,hosein2012,azarbonyad2013} emphasize on selecting top ranked candidates as translations of a query term. According to \cite{donnla2005} selection in such a way could miss some important lower ranked candidates. That is to say adding more candidates to final translation causes the swamping effect \cite{donnla2005} or retrieving irrelevant documents to be exact. Consequently it is tradeoff selecting high ranked translations or adding all candidates. Table~\ref{topn2009} represents Mean Average Precision (MAP) of the runs using different dictionaries. Proper nouns for all resources has been handled by Google's MT\footnote{http://translate.google.com}. Google's descending results prove almost reliable rankings for its translations. As a result, adding more candidates may turn retrieval to different subjects. On the other hand there are some improvements after adding more candidates in Aryanpour and Faraazin in some points. In brief, according to the results, important translations are not necessarily provided in high-ranked translations.\\
As \cite{pirkola98,oard2000} state, defining a set of translations for a query term and treating them as instances of a term, can reach reliable weightings. Table~\ref{pirkola2009} shows the results of the INQUERY retrieval runs based on the Aryanpour dictionary and topics in CLEF 2008 and CLEF 2009. The results are relatively better than top ranked selection approach due to resolving the crucial equivalent problem by incorporating all translations. As an important step prior to the INQUERY retrieval run, the proper nouns has been handled by employing Google's MT. 

\begin{table}[!h]
	\caption{Results of top ranked candidate selection in topics of CLEF.}
\begin{center}
    \begin{tabular}{ c p{1.5cm} p{1.4cm} p{1.4cm} p{1.4cm} p{1.4cm} p{1.4cm} p{1.4cm}}
     \hline\noalign{\smallskip}
     Topic~~~~& top $N$ & $\begin{array}{l}\text{Arya-}\\\text{npour}\end{array}$ &  \%Mono & Google &\%Mono& Faraazin&\%Mono\\
    \noalign{\smallskip}
    \hline
    \noalign{\smallskip}
   \multirow{5}{*}{\rotatebox{90}{2008}}
    & top 1 & \textbf{0.2344}& \textbf{52.7}& \textbf{0.2692}& \textbf{60.5}& 0.2145 & 48.2\\ 
    & top 2& 0.2207& 49.6& 0.2620& 58.9&0.2178 & 48.9\\ 
    & top 3& 0.2254& 50.7& 0.2428& 54.6& \textbf{0.2254} & \textbf{50.7}\\
    & top 5& 0.2242& 50.4& 0.2147& 48.2& 0.2120 & 47.6\\
    & top 10& 0.2170& 48.8& 0.1831& 41.1& 0.1812 & 40.7\\   
    \noalign{\smallskip}
    \hline\hline
    \noalign{\smallskip}
    \multirow{5}{*}{\rotatebox{90}{2009}} &
    top 1     & 0.1942& 47.7&\textbf{0.2559}& \textbf{62.9}& 0.2150&
     52.8\%\\ 
    & top 2& 0.2042& 50.2& 0.2543& 62.5& 0.2199& 
     54.0\%\\ 
    & top 3& 0.2019& 49.6& 0.2460&60.4 & \textbf{0.2216}& 
     \textbf{54.4}\%\\
    & top 5& \textbf{0.2197}&\textbf{54.0}& 0.2248& 55.2& 0.2161&  
     53.1\%\\
    & top 10& 0.2079& 51.1& 0.2166& 53.2& 0.2018& 
     49.6\%\\
    \noalign{}
    \hline
    \end{tabular}
    \label{topn2009}
\end{center}
\end{table}

\vspace{-0.9cm}
\begin{table}[!h]
\caption{Results of Pirkola's structured query algorithm on CLEF.}
\begin{center}
    \begin{tabular}{  l p{1.2cm} p{1.2cm} p{1.2cm} p{1.2cm} p{1.2cm} p{1.2cm} p{1.2cm} p{1.2cm}}
    \hline\noalign{\smallskip}
    Topic~~~~& Measure & $\begin{array}{l}\text{Arya-}\\\text{npour}\end{array}$ & \%Mono& Google & \%Mono& Faraazin & \%Mono& Mono\\
    \noalign{\smallskip}
    \hline
    \noalign{\smallskip}
    2008 & Map     & 0.2706& 60.8& 0.2632& 59.2& 0.2493& 
    56.0& 0.4449\\ 
    & Prec@5  & 0.4920& 69.9& 0.4920&69.9&  0.4360& 
    61.9& 0.7040\\ 
    & Prec@10 & 0.4740& 70.5& 0.4600& 68.5& 0.4280&
    63.7& 0.6720\\
    \noalign{\smallskip}
    \hline\hline
    \noalign{\smallskip}
    2009& Map     & 0.2593& 63.7& 0.2752& 67.6& 0.2328&
    57.2& 0.4070\\ 
    & Prec@5  & 0.4840& 80.7& 0.4760& 79.4& 0.3960&
    66.0& 0.6000\\ 
    & Prec@10 & 0.4160& 69.6& 0.4500& 75.2& 0.3760& 
    62.9& 0.5980\\

    \hline
    \end{tabular}
    \label{pirkola2009}
\end{center}
\end{table}

\subsection{MESC and Probabilistic Selection Model}
\begin{figure}[!h]%
    \centering
    \subfloat[Precision-Recall for CLEF 2008]
    {{\includegraphics[width=0.45\textwidth]{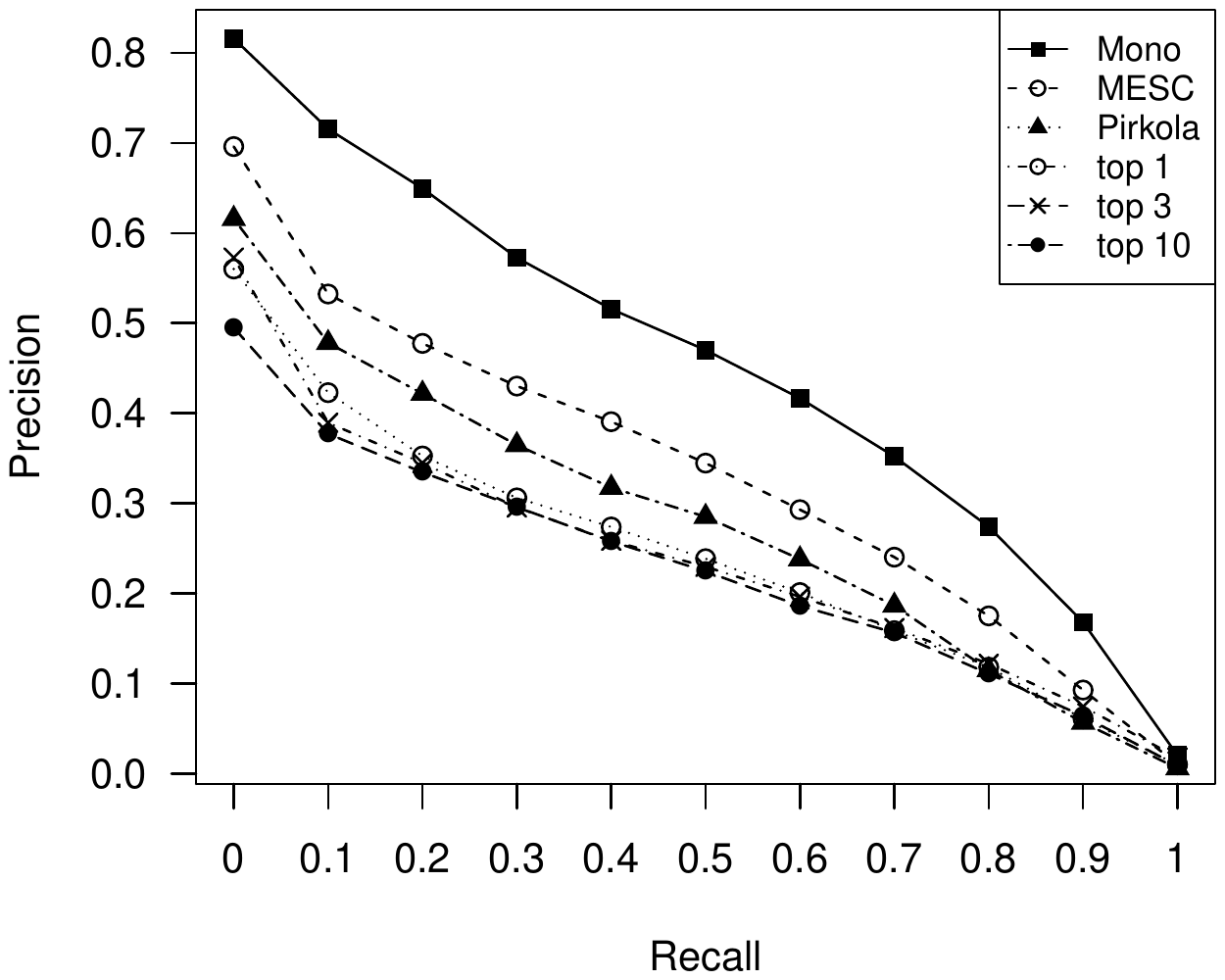} }}%
    \qquad
    \subfloat[Precision-Recall for CLEF 2009]
    {{\includegraphics[width=0.45\textwidth]{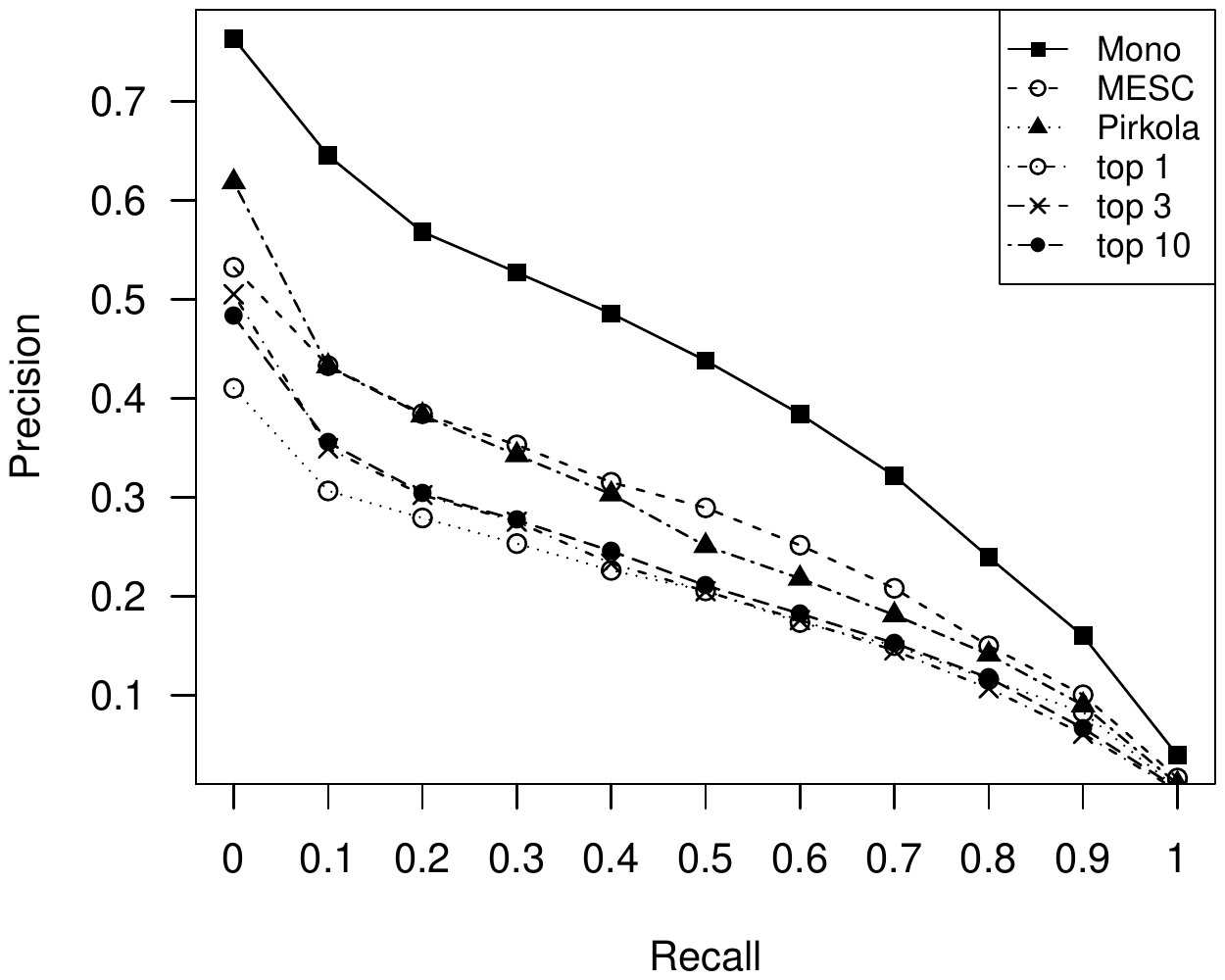} }}%
    \caption{Comparison of MESC performance against previous works by Aryanpour.}%
    \label{fig:PrecisionRecall}%
\end{figure}

As represented in Table~\ref{scaleTable} Aryanpour has longest candidate lists and the results of top ranked selection approach support the view that it is more difficult to select the best candidate in a high coverage dictionary compared to the lower coverage ones. Nevertheless, having high coverage could benefit the MESC algorithm's performance for the following reasons: firstly, neither the number of candidates nor the ranks of them effect the performance of MESC and secondly containing more crucial equivalents benefit MESC. Table~\ref{MESC2009} shows the results of applying MESC on the CLEF dataset. Improvements of MESC compared to the INQUERY retrieval system are also presented. A closer look at the results of Table~\ref{MESC2009} indicates substantial improvements in terms of MAP for all mentioned dictionaries. However in topics of CLEF 2009 there are less improvements comparing to INQUERY. The statistics of the topics in CLEF 2009 in Fig. ~\ref{proportions} show more phrases in CLEF 2008 compared to CLEF 2009. Extracting different formations of a lexicon in phrases is the main superiority of MESC and the Pirkola's structured query approach.\\
Nonetheless we can state missing important translations of a term in a dictionary as another reason to fail accurate translation in MESC. For example in query \textit{`Stress and Health'}, transliterating \textit{`Stress'} is better translation against choosing \textit{`f\textipa{S}$\hat{a}$r'} that means pressure. In such situations, MESC is limited to the provided candidates which may differ in subject with the original query.\\
Fig.~\ref{fig:PrecisionRecall} presents precision at different levels of recall. As a reason for lower precision in some points for CLEF 2009, we can point to synonym equivalent effect. For instance in query \textit{`Tourist Attractions'}, for the query term \textit{`Tourist'}, its transliteration is a correct translation which Aryanpour does not provide. Despite missing such a meaningful candidate, either \textit{`grd\textipa{S}gr'} or \textit{`jh$\hat{a}$ngrd'} is a trustful translation and both of them have significant usages in similar contexts. The structured query approach considers both of them and the INQUERY retrieval system scores documents based on containing either the former translation or the latter one. Under these circumstances, INQERY has better performance. 
We can state multi part translations as another reason for degraded precision. In query \textit{`Freight Transport by Rail'}, the term \textit{`Rail'} means railroad. \textit{`r$\hat{a}$h-$\hat{a}$hn'} is a most common translation for the query term. In Hamshahri \textit{`r$\hat{a}$h'} and \textit{`$\hat{a}$hn'} almost always are separated by a space. Missing effective tokenizers in highly inflected languages like Persian, causes MESC to treat such parts as distinct terms. Finding these parts and merging them by a half space benefit the MESC algorithm to compute reliable probabilities.\\

\vspace{-0.5cm}
\begin{table}[!h]
	\caption{Results of applying MESC algorithm on CLEF.}
\begin{center}
    \begin{tabular}{l p{1.2cm} p{1cm} p{0.8cm} p{0.9cm} p{0.9cm} p{0.9cm} p{0.9cm} p{0.9cm} p{0.8cm} p{0.8cm} p{0.8cm}}
    \hline\noalign{\smallskip}
    Topic~~~~& Measure & $\begin{array}{l}\text{Arya-}\\\text{npour}\end{array}$ & \%M& Impr$_.$& Google& \%M & Impr$_.$& $\begin{array}{l}\text{Fara-}\\\text{zin}\end{array}$ & \%M& Impr$_.$& Mono\\
    \noalign{\smallskip}
    \hline
    \noalign{\smallskip}
    2008 & Map& 0.3215&  72.2& +18.8& 0.319& 71.7& +15.9& 0.293 & 65.8&+25.7& 0.4449\\ 
    & Prec@5  & 0.50& 71.0& +1.6& 0.492& 69.9& +0.0& 0.44 & 62.5&+0.9& 0.704\\ 
    & Prec@10 & 0.502& 74.7& +5.5& 0.49&  72.9& +6.5& 0.436 & 64.9&+1.9&0.672\\
    \noalign{\smallskip}
    \hline \hline
    \noalign{\smallskip}
    2009 & Map     & 0.2682& 65.9& +3.4& 0.278& 68.2& +0.9& 0.247 & 60.7&+6.1& 0.407\\ 
    & Prec@5  & 0.428& 71.3& -11.7& 0.456& 76.0& -4.2& 0.364 & 60.7&-8.0& 0.60\\ 
    & Prec@10 & 0.408& 68.2& -1.9& 0.424& 70.9& -5.8& 0.354 & 59.2&-5.8& 0.598\\
    \hline
    \end{tabular}
    \label{MESC2009}
\end{center}
\end{table}
\vspace{-0.5cm}

\section{Conclusions and Future Works}
\label{Conclusion}
In this research we proposed MESC algorithm, a probabilistic dictionary-based CLIR approach for agglutinative languages. MESC initially provides omitted formations of dictionary candidates using the Levenshtein algorithm and a large monolingual collection. Secondly MESC uses a probabilistic candidate selection model to select most accurate translations. Experimental results of English-Persian CLIR runs provide confirmatory evidences in favor of the MESC algorithm against the Pirkola's INQUERY retrieval system. They also support the view that using a bilingual machine readable dictionary with a reliable coverage improves the CLIR performance regardless of considering any rankings. As an important future work we can state considering phrase detection algorithms and applying the proposed MESC locally to extract truthful support candidates.
\vspace{-0.3cm}
\section{Acknowledgments}
\vspace{-0.3cm}
We would like to thank Behzad Mirzababaei from Natural Language Processing Lab for his helpful comments and Iman D. Behbahani from Intelligent Information Systems Lab for his advantageous advice.

\bibliographystyle{splncs03.bst}
\bibliography{ref}

\begin{thebibliography}{10}
\providecommand{\url}[1]{\texttt{#1}}
\providecommand{\urlprefix}{URL }

\bibitem{hamshahri2009}
AleAhmad, A., Amiri, H., Darrudi, E., Rahgozar, M., Oroumchian, F.: Hamshahri:
  A standard persian text collection. Know.-Based Syst.  22(5),  382--387 (Jul
  2009)

\bibitem{hosein2012}
Azarbonyad, H., Shakery, A., Faili, H.: Using learning to rank approach for
  parallel corpora based cross language information retrieval. In: ECAI'12. pp.
  79--84 (2012)

\bibitem{azarbonyad2013}
Azarbonyad, H., Shakery, A., Faili, H.: Exploiting multiple translation
  resources for english-persian cross language information retrieval. In:
  Information Access Evaluation. Multilinguality, Multimodality, and
  Visualization, pp. 93--99. Springer Berlin Heidelberg (2013)

\bibitem{cao2007}
Cao, G., Gao, J., Nie, J.Y., Bai, J.: Extending query translation to
  cross-language query expansion with markov chain models. In: Proceedings of
  the sixteenth ACM conference on Conference on information and knowledge
  management. pp. 351--360. CIKM '07, ACM, New York, NY, USA (2007)

\bibitem{chen2001}
Chen, A., Jiang, H., Gey, F.: English-chinese cross-language ir using bilingual
  dictionaries (2001)

\bibitem{Ehsan2013}
Ehsan, N., Faili, H.: Grammatical and context-sensitive error correction using
  a statistical machine translation framework. Softw., Pract. Exper.  43(2),
  187--206 (2013)

\bibitem{donnla2005}
Gearailt, D.N., Gearailt, C.D.N., College, C.: Dictionary characteristics in
  cross-language information retrieval (2005)

\bibitem{hashemi2011}
Hashemi, H.: {Using Comparable Corpora for English-Persian Cross-Language
  Information Retrieval}. Master's thesis, University of Tehran, Tehran (2011)

\bibitem{hashemi2014}
Hashemi, H.B., Shakery, A.: Mining a persian-english comparable corpus for
  cross-language information retrieval. Inf. Process. Manage.  50(2),  384--398
  (Mar 2014)

\bibitem{jurafsky2009speech}
Jurafsky, D., Martin, J.: Speech and Language Processing: An Introduction to
  Natural Language Processing, Computational Linguistics, and Speech
  Recognition. Prentice Hall series in artificial intelligence, Pearson
  Prentice Hall (2009)

\bibitem{kishida2005}
Kishida, K.: Technical issues of cross-language information retrieval: a
  review. Inf. Process. Manage.  41(3),  433--455 (May 2005)

\bibitem{Kishida01082009}
Kishida, K., Ishita, E.: Translation disambiguation for cross-language
  information retrieval using context-based translation probability. Journal of
  Information Science  35(4),  481--495 (2009)

\bibitem{levenshtein1966bcc}
Levenshtein, V.: {Binary Codes Capable of Correcting Deletions, Insertions and
  Reversals}. Soviet Physics Doklady  10,  707 (1966)

\bibitem{levow2005}
Levow, G.A., Oard, D.W., Resnik, P.: Dictionary-based techniques for
  cross-language information retrieval. Inf. Process. Manage.  41(3),  523--547
  (May 2005)

\bibitem{liu2005}
Liu, Y., Jin, R., Chai, J.Y.: A maximum coherence model for dictionary-based
  crosslanguage information retrieval. In: In SIGIR ’05: Proceedings of the
  28th annual international ACM SIGIR conference on Research and development in
  information retrieval. pp. 536--543. ACM Press (2005)

\bibitem{Miangah2014}
Miangah, T.M.: Farsispell: A spell-checking system for persian using a large
  monolingual corpus. LLC  29(1),  56--73 (2014)

\bibitem{nie2010}
Nie, J.Y.: Cross-language information retrieval. Synthesis Lectures on Human
  Language Technologies  3(1),  1--125 (2010)

\bibitem{oard2000}
Oard, D.W., Wang, J.: Ntcir-2 ecir experiments at maryland: Comparing pirkola's
  structured queries and balanced translation. In: In Second National Institute
  of Informatics (NII) Test Collection Information Retrieval (NTCIR) workshop.
  forthcoming (2000)

\bibitem{pirkola98}
Pirkola, A.: The effects of query structure and dictionary setups in
  dictionary-based cross-language information retrieval. In: Proceedings of the
  21st annual international ACM SIGIR conference on Research and development in
  information retrieval. pp. 55--63. SIGIR '98, ACM, New York, NY, USA (1998)

\bibitem{pirkola2001}
Pirkola, A., Hedlund, T., Keskustalo, H., Järvelin, K.: Dictionary-based
  cross-language information retrieval: Problems, methods, and research
  findings. Information Retrieval  4(3-4),  209--230 (2001)

\bibitem{Robertson94}
Robertson, S.E., Walker, S.: Some simple effective approximations to the
  2-poisson model for probabilistic weighted retrieval. pp. 232--241. SIGIR '94
  (1994)

\bibitem{Stolcke02}
Stolcke, A.: Srilm - an extensible language modeling toolkit. In: Hansen,
  J.H.L., Pellom, B.L. (eds.) INTERSPEECH. ISCA (2002)

\bibitem{young:1994}
Young, P.G.: Cross-Language Information Retrieval Using Latent Semantic
  Indexing. Master-thesis, University of Tennessee, Knoxville, Knoxville (1994)

\bibitem{Zhou:2008:HTE:1362782.1362784}
Zhou, D., Truran, M., Brailsford, T., Ashman, H.: A hybrid technique for
  english-chinese cross language information retrieval  7(2),  5:1--5:35 (Apr
  2008)

\end{thebibliography}

\end{document}